\documentclass[prl,showpacs,twocolumn,superscriptaddress]{revtex4}

\usepackage{amsmath}
\usepackage{graphicx}

\usepackage{amsmath}
\usepackage{amssymb}
\usepackage{graphicx}

\begin{document}

\title{Subdiffusion and weak ergodicity breaking in the presence of a reactive
boundary}

\author{Michael A. Lomholt}
\affiliation{Physics Department, University of Ottawa, Pavillon
MacDonald, Ottawa, Ontario K1N 6N5, Canada}
\author{Irwin M. Zaid}
\affiliation{Physics Department, Carleton University, Herzberg Building,
Ottawa, Ontario K1S 5B6, Canada}
\author{Ralf Metzler}
\affiliation{Physics Department, Technical University of Munich, 85748
Garching, Germany}
\affiliation{Physics Department, University of Ottawa, Pavillon
MacDonald, Ottawa, Ontario K1N 6N5, Canada}


\begin{abstract}
We derive the boundary condition for a subdiffusive particle interacting
with a reactive boundary with finite reaction rate.
Molecular crowding conditions, that are found to cause subdiffusion of
larger molecules in biological cells, are shown to effect long-tailed
distributions with identical exponent for both the unbinding times from
the boundary to the bulk and the rebinding times from the bulk. This
causes a weak ergodicity breaking: typically, an individual particle
either stays bound or remains in the bulk for
very long times. We discuss why this may be beneficial for
\emph{in vivo\/} gene regulation by DNA-binding proteins,
whose typical
concentrations are nanomolar. %(Ruusala and Crothers, PNAS 89 (1992) 4903)
\end{abstract}

\maketitle

The interaction of a diffusive particle with a reactive boundary is of
fundamental importance in interface science and technology, e.g., to
transport in porous media \cite{levitzstapf},
interactions of proteins with artificial surfaces and membranes
\cite{proteins}, or applications such as foam relaxation and surfactants
\cite{surf}. For a Brownian particle this has been studied
extensively, expecially concerning the question how bulk exchange
influences the surface distribution of intermittently adsorbed
particles \cite{bychuk}. Here, we derive the exchange dynamics with a
reactive boundary of a subdiffusing particle, whose unbinding
and rebinding times in a molecular crowding environment are both
shown to follow long-tailed distributions.
We demonstrate a weak ergodicity breaking for the particle
trajectory.

This is of particular interest for the search of DNA-binding proteins for
their specific binding site on DNA involving successive events of non-specific
binding to the DNA and bulk excursions, such that the time spent in either of
these events is important in the understanding of the various stochastic
mechanisms involved in (bacterial) gene regulation \cite{bvh,slutsky,michael}.
While the generally applied assumption of Brownian diffusion of
proteins works well for typical \emph{in vitro\/} experiments under dilute
conditions, \emph{in vivo\/} the abundance of a multitude of
biomacromolecules in the cellular cytoplasm have been shown to cause a state
of \emph{molecular crowding}: large molecules such as
proteins, lipids, RNA molecules and ribosomes make up up to 40\% of the
cytoplasmic volume \cite{ellis,takahashi}. In this superdense environment
they hinder each other's motion, causing \emph{subdiffusion\/}
\cite{golding,weiss,banks,weiss1}, with a
mean squared displacement $\langle\mathbf{r}^2(t)\rangle
\propto t^{\alpha}$; $0<\alpha<1$ being a dynamic exponent \cite{report}.
By fluorescent methods, subdiffusion was verified for proteins in membranes
with $\alpha=0.7$ \cite{weiss1}, for proteins in a molecular crowded \emph{in
vitro\/} environment with $\alpha=0.75$ at higher densities \cite{banks},
as well as in the cytoskeleton \emph{in vivo\/} for
messenger RNA of physical size $\approx 100$nm with $\alpha\approx 0.75$
\cite{golding} and for dextran molecules ranging from 10kD to 2 MD with
$\alpha$ in between 0.59 and 0.84 \cite{weiss}. The occurrence of subdiffusion
for particles with mass as low as 10kD was also confirmed by computer
simulations \cite{weiss}. The Lac repressor, a typical DNA-binding protein,
has 141kD \cite{chakerian}, for which a corresponding
$\alpha\approx0.73$ was found \cite{weiss}. Thus, under molecular
crowding conditions, $\alpha\approx0.75$ seems a fairly standard value
for DNA-binding proteins and larger polynucleotides \cite{banks,golding,weiss}.
The time scale over which this subdiffusion persists is not known precisely,
but appears to be longer than minutes, so that
the following considerations are expected to be relevant for genetic
processes \cite{golding,weiss,weiss1,banks}.

To derive the generalized reactive boundary condition, we pursue a continuous
time random walk approach similar to Ref.~\cite{sokolov06}: A
subdiffusing particle jumps from one point to the next after a
waiting time distributed according to the long-tailed probability density
$\psi(t)\simeq\tau^{\alpha}/t^{1+\alpha}$ ($0<\alpha<1$) \cite{scher}. We
start our derivation with the one-dimensional lattice, on which $A_i$ is the
probability to find the particle at lattice point $i=1,2,3,\dots$. The
probability of being at the reactive site (lattice point next to the
boundary) is $\mathcal{A}_0$, the notation indicating that at site 0
the particle can be exchanged with the bound state with rate $\kappa$.
The balance equations then read
\begin{subequations}
\label{eq:govA}
\begin{eqnarray}
&&dA_i(t)/dt=I^+_i(t)-I^-_i(t),\\
&&d\mathcal{A}_0(t)/dt=I^+_0(t)-I^-_0(t)-\kappa\mathcal{A}_0(t),
\label{eq:govA0}
\end{eqnarray}
\end{subequations}
and the loss from a given lattice site due to diffusion is
\begin{subequations}
\label{losses}
\begin{eqnarray}
I^-_i(t)&=&\psi(t)A_i(0)+\int_0^t\;\psi(t-t')I^+_i(t')dt'\\
%&\equiv&\int_0^t\;\Phi_0(t-t')A_i(t') dt',\quad i>0\;,\\
I^-_0(t)&=&\psi_{\kappa}(t)\mathcal{A}_0(0)+\int_0^t\;\psi_{\kappa}(t-t')
I^+_0(t')dt',
%\\
%&\equiv&\int_0^t\;\Phi_\kappa(t-t')A_0'(t')dt'.
\end{eqnarray}
\end{subequations}
where $\psi_{\kappa}(t)\equiv\psi(t)e^{-\kappa t}$. Substituting for $I^+$
from Eqs.~(\ref{eq:govA}), we rephrase Eqs.~(\ref{losses}) in the form
$I^-_i(t)=\int_0^t\Phi(t-t')A_i(t')dt'.$
The kernel $\Phi(t)$ is defined by $\Phi(u)=u\psi(u)/\left[1-\psi
(u)\right]$ in the Laplace domain, $\Phi(u)=\int_0^{\infty}\Phi(t)e^{-ut}dt$
\cite{REMMMM}.
An analogous relation holds for $I^-_0(t)$, with the kernel $\Phi_{\kappa}(u)=
\Phi(u+\kappa)$. For the gain to site $i$ we have, assuming
that the particle jumps to left and right equally likely,
$I^+_i=I^-_{i-1}/2+I^-_{i+1}/2$ and $I^+_0=I^-_{0}/2+I^-_{1}/2$.
Note that if the particle attempts to jump left from site 0, it will be
returned back to the same site.

For the continuum limit, we introduce a new time-dependent
quantity $A_0(t)$ by
$\int_0^t\Phi(t-t')A_0(t')dt'\equiv\int_0^t\Phi_{\kappa}(t-t')
\mathcal{A}_0(t')dt'$,
corresponding to $\mathcal{A}_0(u)=\Phi(u)A_0(u)/\Phi_{\kappa}(u)$.
Combining above results, we find
\begin{equation}
\frac{d A_i(t)}{dt}=\int_0^t\;\Phi(t-t')\frac{A_{i-1}(t')+A_{i+1}(t')-
2A_i(t')}{2}dt'.
\end{equation}
In the continuum limit $A(x=ai,t)=A_i(t)/a$ with the lattice spacing $a$
\cite{REM}, this equation for $i\ge1$ yields
\begin{equation}
\label{eq:subeq}
\frac{\partial A(x,t)}{\partial t}=\frac{a^2}{2}\int_0^t\;\Phi(t-t')
\frac{\partial^2 A(x,t')}{\partial x^2}dt'.
\end{equation}
In the long time limit $u\tau\ll1$, $\psi(u)\sim1-(u\tau)^{
\alpha}$, and $\Phi(u)\sim u^{1-\alpha}\tau^{-\alpha}$ to leading order.
With $K_{\alpha}=a^2/[2\tau^{\alpha}]$ and the fractional Riemann-Liouville
operator,
\begin{equation}
_0D_t^{1-\alpha}A(x,t)=\frac{1}{\Gamma(\alpha)}\frac{\partial}{\partial t}
\int_0^t\frac{A(x,t')}{(t-t')^{1-\alpha}}dt',
\end{equation}
Eq.~(\ref{eq:subeq}) is equivalent to the fractional diffusion equation
$\partial A(x,t)/\partial t=K_{\alpha}\,_0D_t^{1-\alpha}\partial^2A(x,t)/
\partial x^2$ for $x>0$ \cite{report}.
Similarly, Eq.~(\ref{eq:govA0}) can be recast into the form
\begin{eqnarray}
\nonumber
&&\frac{d}{d t}\mathcal{A}_0(t)+\kappa \int_0^t\;\left(\Phi_\kappa^{-1}\Phi
\right)(t-t')A_0(t')dt'\\
&&\hspace*{0.8cm}=\int_0^t\;\Phi(t-t')\frac{A_1(t')-A_0(t')}{2}d t'.
\end{eqnarray}
In the continuum limit, we recover the expression
\begin{eqnarray}
\nonumber
&&-\delta(t)\mathcal{A}_0(0)+\int_0^t\;\Psi(t-t')A(0,t')dt'\\
&&\hspace*{0.8cm}
=\frac{a^2}{2}\int_0^t\;\Phi(t-t')\left.\frac{\partial A(x,t')}{
\partial x}\right|_{x=0}dt'
\label{eq:subbound}
\end{eqnarray}
with $\Psi(u)=a(u+\kappa)\Phi(u)/\Phi_\kappa(u)$. $\mathcal{A}
_0(0)$ is 1, if the particle is initially released at site 0, and 0
otherwise. The reaction rate at the boundary is $j_{\rm react}=a\kappa\int_0
^t\;\left(\Phi_\kappa^{-1}\Phi\right)(t-t')A(0,t')dt'$,
and the right hand side of Eq.~(\ref{eq:subbound}) represents the
flux into $x=0$ from positive $x$. 
We expand Eq.~(\ref{eq:subbound}) at $u=0$ in Laplace space (note that
$u\ll \kappa$ \cite{REM}) producing the sought for reactive boundary condition
\begin{equation}
\label{eq:subbound2}
K_\alpha\,{}_0D_t^{-\alpha}\left.\frac{\partial A(x,t)}{\partial x}\right|_{
x=0}=-\mathcal{A}_0(0)+k\,_0D_t^{-\alpha}A(0,t)
\end{equation}
for the subdiffusive particle. This is one of the main results of this work.
We defined $k=2\kappa K_\alpha/\left[a\Phi_\kappa(u=0)\right]\sim a\kappa^
\alpha$, using that $\kappa\tau\to 0$ in the continuum limit \cite{REM}.

The Berg-von Hippel model maps the
binding/unbinding dynamics of a DNA-binding protein to/from the DNA
surface onto
a cylinder of radius $R_1$ placed along the $z$-axis in cylindrical
coordinates $(r,\theta,z)$ \cite{bvh}. For particles subdiffusing in the
space $r>R_1$ with density $P$ the boundary condition (\ref{eq:subbound2})
generalizes to
\begin{eqnarray}
\nonumber
&&2\pi R_1K_\alpha\, {}_0D_t^{-\alpha}\left.\partial P(r,t)/\partial
r\right|_{r=R_1}\\
&&\hspace*{0.8cm}=-P_0/L+ k_{\rm on}\,{}_0 D_t^{-\alpha}\left.
P\right|_{r=R_1},
\end{eqnarray}
where $k_{\rm on}=2\pi R_1 k$, and we have assumed rotational and translational
symmetry around and along the $z$-axis. $L$ is the length of the cylinder along
the $z$-axis, and $P_0=1$ if at $t=0$ the particle is at the boundary, and
$P_0=0$
otherwise. The rate of reaction with the cylinder per length along the $z$-axis
is $j_{\rm react}(t)=k_{\rm on}\,{}_0 D_t^{1-\alpha}\left.P\right|_{r=R_1}$,
and the fractional diffusion equation (\ref{eq:subeq}) is replaced by
\begin{equation}
\label{fd1}
\frac{\partial P}{\partial t}=K_\alpha\, {}_0 D_t^{1-\alpha}\frac{1}{r}
\frac{\partial}{\partial r}\left(r\frac{\partial P}{\partial r}\right).
\end{equation}

Consider now the situation when the particle is
\emph{bound\/} at the boundary at $t=0$. In a crowded environment the full
escape of the particle from the boundary consists of two steps: unbinding
with rate $\kappa_{\rm off}$, returning it to the exchange site 0, and then
avoiding to rebind to the boundary such that the particle is unbound
at the moment when the environment allows for jumping to site 1. If the
particle is bound at the time when the environment allows a jump (in
principle), then the process needs to start over, etc. Thus, the waiting
time distribution for the full escape can be written as
\begin{eqnarray}
\nonumber
&&\wp_{\mathrm{unb}}(t)=R(t)+\int_0^tR(t-t')\Pi(t')dt'\\
&&+\int_0^tR(t-t')\int_0^{t'}\Pi(t'-t'')\Pi(t'')dt''dt'+\ldots,\qquad
\end{eqnarray}
with $R(t)=\tilde{\psi}(t)[1-\mathrm{P}_{\mathrm{bound}}(t)]$ and $\Pi(t)=
\tilde{\psi}(t)\mathrm{P}_{\mathrm{bound}}(t)$. Here $\mathrm{P}_{\mathrm{
bound}}(t)$ is the probability that the particle is bound at time $t$, given
that it was bound at $t=0$ and that the environment has not yet opened up to
allow a jump to $i=1$. This yields
\begin{equation}
\mathrm{P}_{\mathrm{bound}}(t)=\frac{\kappa}{\kappa+\kappa_{\rm off}}
+\frac{\kappa_{\rm off}}{\kappa+\kappa_{\rm off}}e^{-(\kappa+\kappa_{\rm
off})t}.
\end{equation}
The waiting time $\tilde{\psi}(t)$ between opening events in the environment,
\emph{allowing\/} a jump to $i=1$, is given by
\begin{eqnarray}
\nonumber
\tilde{\psi}(t)&=&\frac{\psi(t)}{2}+\int_0^t\frac{\psi(t-t')}{2}\frac{
\psi(t')}{2}dt'\\
&&\hspace*{-1.4cm}
+\int_0^t\frac{\psi(t-t')}{2}\int_0^{t'}\frac{\psi(t'-t'')}{2}
\frac{\psi(t'')}{2}dt'dt''+\dots,
\end{eqnarray}
where the factor 1/2 accounts for the fact that the particle jumps to site
1 only with a probability 1/2 (and stays at 0 with probability 1/2).
In Laplace space, $\tilde{\psi}(t)$ can be expressed in closed form through a
geometric series,
$\tilde{\psi}(u)=[\psi(u)/2]/[1-\psi(u)/2]\sim 1-2(u\tau)^\alpha$.
Similarly, in the Laplace domain,
\begin{eqnarray}
\nonumber
\wp_{\mathrm{unb}}(u)&=&{\tilde \psi}(u)-\Pi(u)+[\tilde{\psi}(u)-\Pi(u)]
\Pi(u)\\
&&\hspace*{-1.6cm}+[{\tilde \psi}(u)-\Pi(u)]\Pi(u)^2+\dots
=\frac{{\tilde \psi}(u)-\Pi(u)}{1-\Pi(u)},
\end{eqnarray}
where $\Pi(t)$ in Laplace space assumes the exact form
\begin{equation}
\Pi(u)=\frac{\kappa}{\kappa+\kappa_{\rm off}}{\tilde \psi}(u)+\frac{\kappa_{\rm
off}}{\kappa+\kappa_{\rm off}}{\tilde \psi}(u+\kappa+\kappa_{\rm off}).
\end{equation}
Collecting the results, we obtain for $\wp_{\mathrm{unb}}$ at small $u$
\cite{REM}
\begin{equation}
\wp_{\mathrm{unb}}(u)\sim 1-\frac{\kappa+\kappa_{\rm off}}{\kappa_{\rm off}(
\kappa+\kappa_{\rm off})^\alpha}u^\alpha=1-u^\alpha/k_{\rm off},
\end{equation}
where in the continuum limit ($\kappa_{\mathrm{off}}\ll\kappa$)
we have $k_{\rm off}\sim\kappa_{\mathrm{off}}/\kappa^{1-\alpha}$.
The unbinding times then are distributed according to the
power-law $\wp_{\mathrm{unb}}\simeq 1/\left(k_{\rm off}t^{1+\alpha}
\right)$. This is a central finding of this work: The crowded
environment impeding the desorption to the bulk translates the a priori
exponential distribution of unbinding times to a power-law \cite{REMM}.
Once arrived at site $i=1$, the particle subdiffuses in the
bulk. We consider here the cylindrical case governed by Eq.~(\ref{fd1}).
With initial condition $P_0=1$ and
a reflecting boundary condition at $r=R_2$ \cite{REMMM},
an analytic result can be obtained in terms of
modified Bessel functions, see Ref.~\cite{zaid} for
details. A systematic expansion for small $u$ leads to the result
\cite{zaid}
\begin{equation}
\wp_{\mathrm{reb}}(u)\sim 1-Su^\alpha/k_{\mathrm{on}}.
\end{equation}
with the cylindrical cross-section $S=\pi(R_2^2-R^2_1)$. The form
$\wp_{\mathrm{reb}}(t)\sim t^{-1-\alpha}$ is typical for subdiffusion
\cite{report}.

\begin{figure}
\includegraphics[width=7.2cm]{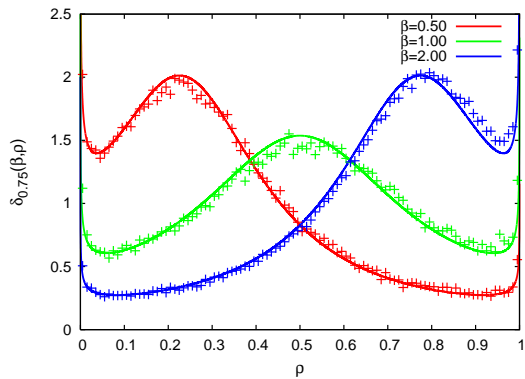}
\caption{The distribution $\delta_{\alpha}$, Eq.~(\ref{lamperti}), for
various $\beta$, with
$\alpha=0.75$. In all cases, a divergence at $\rho=0$ and 1 is observed.
The points are results from a stochastic simulation \cite{zaid}.}
\label{delta2}
\end{figure}

Both unbinding to the volume and returning
to the reactive boundary follow power-law forms with identical
asymptotic behavior $\sim t^{-1-\alpha}$. The lack of a characteristic
time scale separating micro- and macroscopic events
gives rise to weak ergodicity breaking
\cite{bouchaud1}. As shown in Ref.~\cite{bel06}, the time-averaged probability
in the bound state
$\overline{p}_{\mathrm{bound}}=\lim_{t\to\infty}t_{\mathrm{bound}}/t$
for a \emph{single\/} trajectory, $\overline{p}_{\rm bound}$ has the
distribution
$\mathcal{P}\left(\overline{p}_{\mathrm{bound}}\right)=
\delta_\alpha\Big(k_{\rm on}/(Sk_{\mathrm{off}}),\overline{p}_
{\mathrm{bound}}\Big),$
with the Lamperti-generalized $\delta$-function \cite{lamperti,bel06}
\begin{equation}
\label{lamperti}
\delta_{\alpha}(\beta,\rho)=\frac{\pi^{-1}\sin(\pi\alpha)\beta\rho^{
\alpha-1}(1-\rho)^{\alpha-1}}{\beta^2(1-\rho)^{2\alpha}+\rho^{2\alpha}+2
\beta(1-\rho)^{\alpha}\rho^{\alpha}\cos\pi\alpha}.
\end{equation}
Note that $\mathcal{P}$ is normalized, $\int_0^1\mathcal{P}(\overline{p}_{
\mathrm{bound}})d\overline{p}_{\mathrm{bound}}=1$, and valid
in the long $t$ limit. It is independent of $t$ and in that sense an
equilibrium is attained. However, while in the Brownian limit $\alpha=1$,
ergodicity and a sharply peaked behavior for $\mathcal{P}$ are recovered,
the very distinct behavior of $\mathcal{P}$ for $\alpha<1$
is displayed in Fig.~\ref{delta2} for $\alpha=0.75$: as function of $\rho=
\overline{p}_{\mathrm{bound}}$, the distribution peaks at 0 and 1, with a
smaller maximum in between. Thus, in a single trajectory a particle
is typically either bound or unbound, independently of the duration
of the trajectory. This nonergodic behavior is imposed
on the system by the probability $\int_t^{\infty}\psi(t')dt'\sim t^{
-\alpha}$ of never moving, that decays very slowly. The smaller the cross
section $S$, the more likely is it to find the particle in a bound state,
as it should be. The behavior of $\mathcal{P}$ therefore contrasts the
ensemble average over many trajectories,
$\langle {\overline{p}}_{\rm bound}\rangle=\left(1+Sk_{\mathrm{off}}/
k_{\rm on}\right)^{-1}$, corresponding to the form $\mathcal{P}(\overline{p}_
{\mathrm{bound}})=\delta\left(\overline{p}_{\mathrm{bound}}-k_{\rm on}/[
k_{\rm on}+Sk_{\rm off}]\right)$ \cite{bel06}. This can be understood as
follows.
For an ensemble of particles, $k_{\mathrm{on}}/k_{\mathrm{off}}$ defines the
nonspecific binding constant $K_{\mathrm{ns}}$, equal to the
ratio $N_{\mathrm{bound}}/(SN_{\mathrm{unbound}})$ of bound and unbound
particles normalized by the cross section \cite{michael}.
Then $1/(1+Sk_{\mathrm{off}}/k_{\mathrm{on}})$ is the ensemble probability
that a particle is bound. Weak ergodicity breaking is thus relevant for
systems with few particles of a given species.

Transcription factors (TFs), DNA-binding proteins regulating the
transcription of a specific gene, occur at very small numbers (a few to some
hundred \emph{per cell\/} \cite{guptasarma}), and in many cases it is essential
for the stability of genetic circuits that a TF is always bound at some
operator site on the DNA \cite{audun,ptashne}. While the random motion of the
TFs in
most \emph{in vitro\/} experiments is Brownian, molecular crowding \emph{in
vivo\/} causes subdiffusion of TFs. This would have interesting consequences
for gene regulation. Namely, due to the weakly ergodic behavior demonstrated
here, TFs will typically stay close to their binding site with a diverging
characteristic time scale, such that unbinding and escape to the volume is
greatly reduced. The price to pay is that once a TF
escapes to the bulk, its return is also affected by an infinite average time.
Moreover, there exists a large class of TFs, such as the well-studied Lac and
bacteriophage $\lambda$ repressors in \emph{E.~coli} \cite{ptashne}, whose
specific binding site is located immediately adjacent to their coding region.
Biochemical production occurs likely within a few tens of nm from the coding
region \cite{miller}, and therefore from the targeted binding site. The weak
ergodicity breaking thus keeps those TFs within a small volume around their
complete biochemical cycle, very likely leading to a significant increase
in the stability of the regulation of that particular gene. Subdiffusion
caused by molecular crowding could therefore be very beneficial for
living cells, allowing them to maintain the concentrations of even vital
TFs at nanomolar levels. This may significantly impact our current picture
of gene regulation \emph{in vivo\/} and pose the need to perform experiments
much closer to the cellular crowding conditions in order to obtain meaningful
information for the \emph{in vivo\/} situation.

We derived the generalized reactive boundary condition for the interaction
of a subdiffusive particle with a boundary
and showed that in the molecular crowding scenario the distribution
of unbinding times becomes long-tailed, with the same exponent as the
distribution of return times to the boundary. This gives rise to
weak ergodicity breaking, relevant for systems with small numbers
of diffusing particles. Apart from gene regulation, these effects will impact
cellular processes in more general, such as the interactions of biopolymers
with membrane proteins, or the exchange of shorter DNA and RNA chains across
cellular membranes. Moreover, they will affect trapping phenomena in the
vicinity of soft interfaces in more general, e.g., the
exchange dynamics from ion clouds in the vicinity of charged or polarized
membranes.
It should be very interesting to explore these effects by single particle
tracking under molecular crowding conditions using fluorescent labelling
techniques.

We thank Igor Sokolov and Eli Barkai for helpful discussions, and
acknowledge
funding by NSERC of Canada and the Canada Research Chairs programme.


\begin{thebibliography}{99}
\bibitem{REMMMM} We denote the Laplace transform $f(u)$ of a function $f(t)$
by explicit dependence on the argument.
\bibitem{REM} We keep $K_{\alpha}=a^2/(2\tau^{\alpha})$, $k$, and $k_{\mathrm{
off}}$ fixed, implying $\tau
\simeq a^{2/\alpha}$, $\kappa\simeq a^{-1/\alpha}$, and $\kappa_{\mathrm{off}}
\simeq a^{1-1/\alpha}$.
\bibitem{REMM} A different scenario corresponding to exponential unbinding
in a trapping environment is discussed in Ref.~\cite{zaid}.
\end{thebibliography}
\end{document}